\documentclass{elsart}
\usepackage{graphics}

\def\cf{{\it cf.}}
\def\ltw{\>\hbox{\lower.25em\hbox{$\buildrel <\over\sim$}}\>}
\def\gtw{\>\hbox{\lower.25em\hbox{$\buildrel >\over\sim$}}\>}
\begin{document}
 
\begin{frontmatter}
\title{On Dynamical Models for  Radio Galaxies}


\author[1]{Jean Eilek,}
\author[2]{Phil Hardee,}
\author[1]{Tomislav Markovic,}
\author[4]{Mike Ledlow,}
\author[5]{Frazer Owen}
\address[1]{Astrophysics Research Center, Physics Department, New
Mexico Tech, Socorro, NM 87801}
\address[2]{Physics \& Astronomy Department, University of Alabama,
Tuscaloosa AL 35487}
\address[4]{Institute for Astrophysics, Physics \& Astronomy
Department, University of New Mexico, Albuquerque NM 87131}
\address[5]{National Radio Astronomy Observatory, Socorro, NM
87801\thanksref{Somewhere}} 

\thanks[Somewhere]{The National Radio Astronomy Observatory is
operated by Associated Universities, Inc., under contract with the
National Science Foundation.}

\begin{abstract}

The tailed radio galaxies that have been called ``Type I'' are not a
uniform set.  To study their dynamics, we have used the Ledlow-Owen
data set, which provides a new sample of 250 radio galaxies in nearby
Abell clusters.  These sources divide 
into two clear categories based on their radio morphology.  Type A
sources (``straight'') contain 
nearly straight jets which are embedded in outer radio lobe. Type B
sources (``tailed'') have a well-collimated jet flow which undergoes a
sudden transition, at an inner hot spot, to a less collimated
flow which continues on and forms a radio tail.  We have not found any
separation of these classes in terms of radio power, radio flux size,
galaxy power or external gas density.  We propose the difference is
due to the development, or not, of a disruptive flow instability, such
as Kelvin-Helmholtz, and the saturation of the instability when it
develops. 

\end{abstract}

\end{frontmatter}

\section{Introduction}

We do not yet fully understand the dynamics and
evolution of radio galaxies.  Classifying sources on the basis of
their morphology has been one useful approach to this
question.  Such work by Fanaroff \& Riley
(``FR''; \cite{FR}) has stimulated discussion and simple
models of the sources.  FR Type II sources are generally accepted to
arise from a  collimated, supersonic jet which impacts the ambient medium,
resulting in a hot spot at the end of the jet, and leaving behind
 a cocoon which surrounds the jet.  This model is 
qualitatively attractive and leads to possible quantitative
predictions.   Our picture of FR Type I sources, however, is much less
complete.  It has been suggested that Type I's are subsonic, turbulent
plumes.  This model does not, however, seem able to account for the
range of phenomena seen in the data; nor has it been developed
quantitatively to a point which would allow specific predictions
of source evolution.  We find this frustrating, as Type I sources are
much more common  in the universe than Type II's, and show much more
diversity.   

We present two figures which illustrate how rare 
classical Type II's are.  First,  one of us \cite{JE.LF} has
used the LRL, B2 and GB samples \cite{LRL,B2,MC} to  calculate 1.4
GHz luminosity 
functions separately for Type II and Type I  sources, shown in Figure
1.  It is apparent that Type I's
dominate Type II's below $P_{1.4} \sim  10^{32}$W/Hz, and strongly
dominate in total numbers (integrated over radio flux). 
Second, only 13 of 197 resolved sources in a new data set are classic
II's.  This data set is the Ledlow-Owen sample 
of radio sources in clusters of galaxies (shown in Figure 2;
\cite{LOrad}).  Our work in this paper uses this sample.

\begin{figure}[htb]
\begin{minipage}{0.48\textwidth}
 \resizebox{\textwidth}{!}{\includegraphics{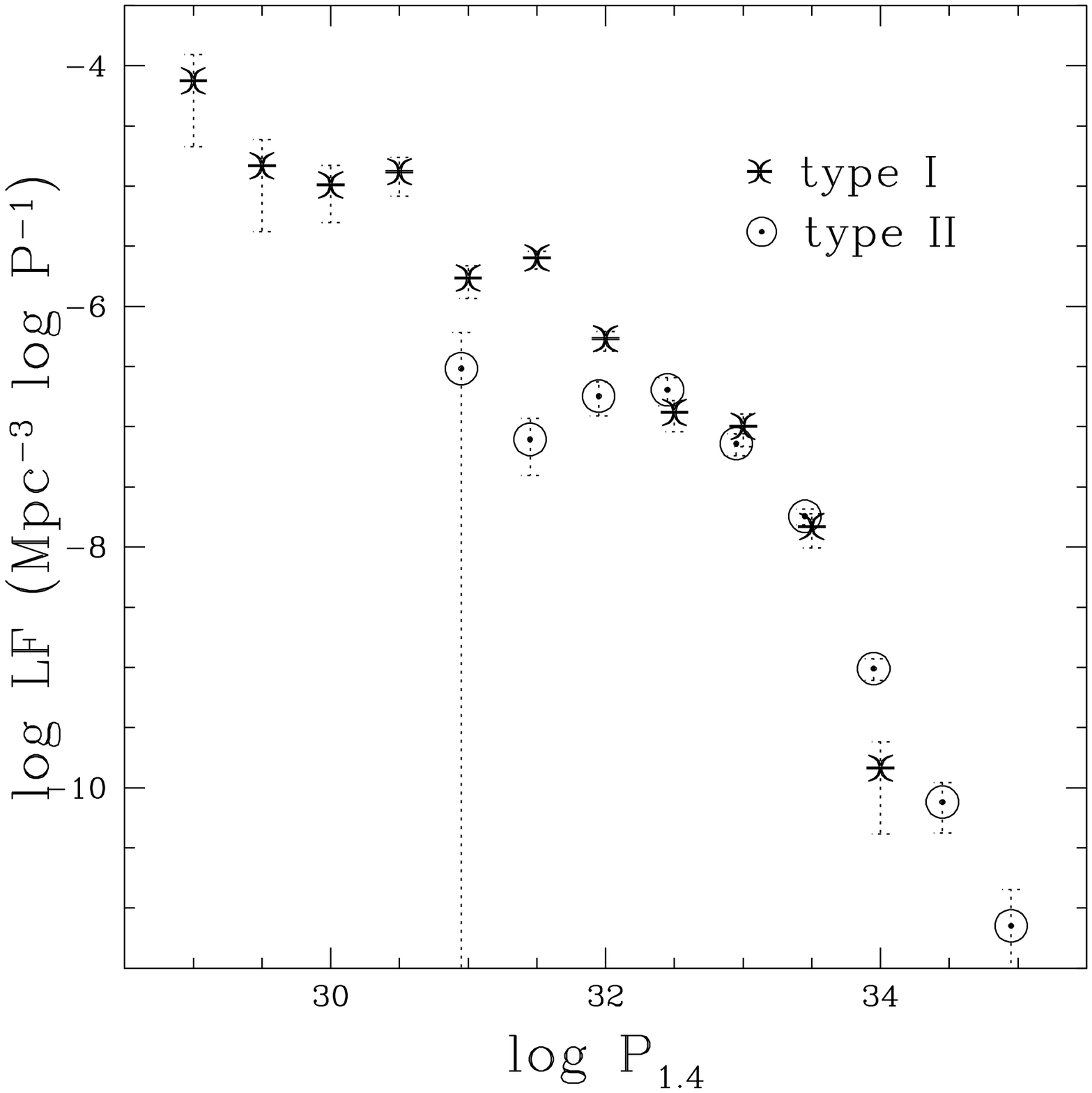}}
\caption[]{Luminosity functions for Type I and Type II sources
separately.  The sources are nearby ($z < 0.5$) members of the 3CR, B2
and GB samples, identified as Type I or Type II from published radio
maps.  The LF's were calculated by the inverse volume method, taking
radio and (for B2) optical flux limits into account. From \cite{JE.LF}.}
\end{minipage}
\hspace{0.04\textwidth}
\begin{minipage}{0.48\textwidth}
\resizebox{\textwidth}{!}{\includegraphics{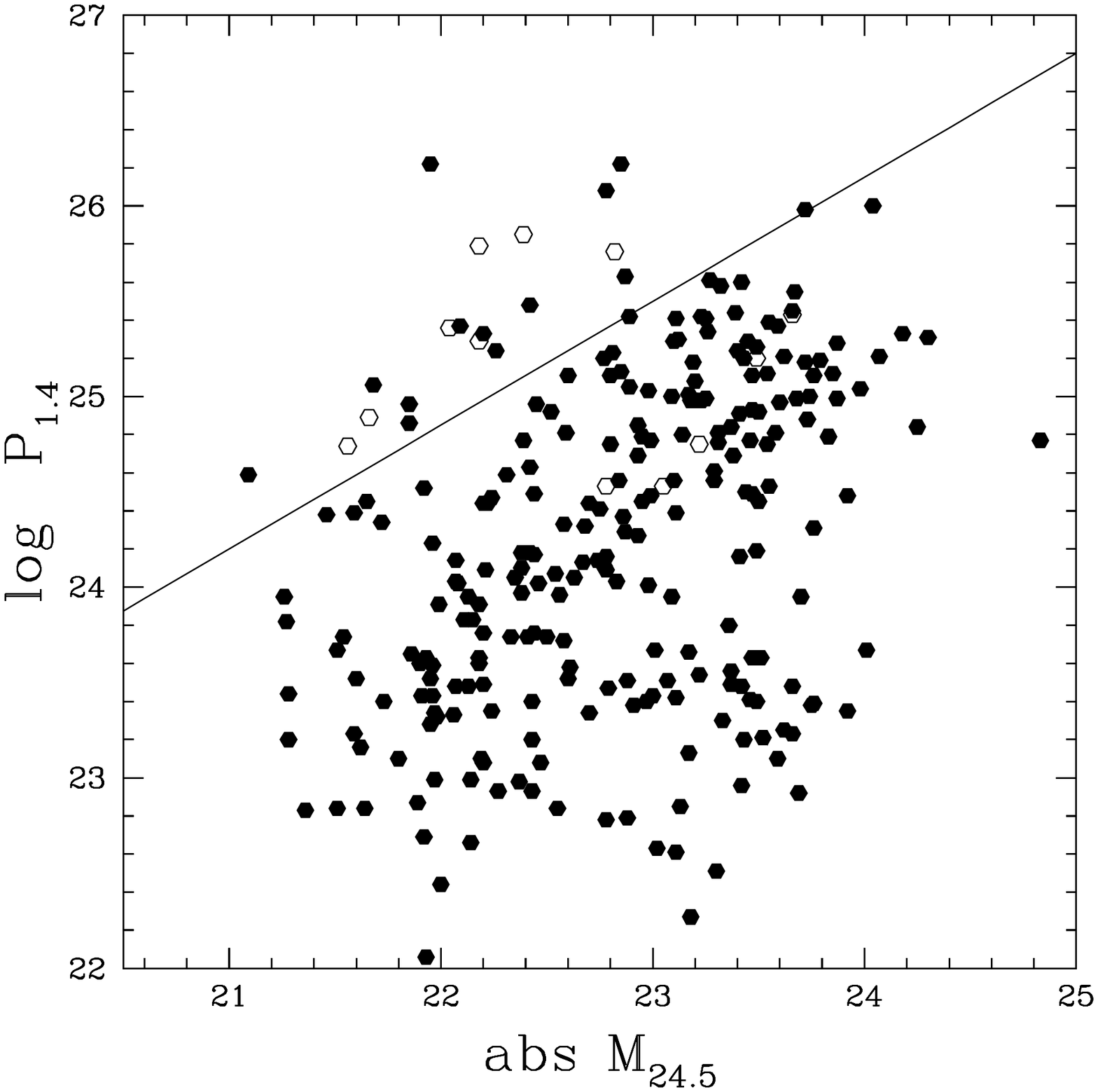}}
\caption[]{The full
Owen-Ledlow sample in the (radio power, optical magnitude) plane, with
the dividing line from \cite{LObreak} shown.  Nearly all of the sample
are in the ``Type I'' region, below the dividing line. We find that
these sources should be further divided into
``Straight'' and ``Tailed'' classes.   Classic Type
II's are shown as open circles.} 
\end{minipage}
\end{figure}

What questions would we like to answer regarding the dynamics of radio
sources?  Many spring to mind:

\begin{itemize}

\item How does a radio source evolve in linear size, volume, morphology?

\item How does the luminosity of a radio source evolve with time?

\item How can we explain the observed distribution of sources in the
(size, power) or (power, galaxy magnitude) planes?

\item What physical factors determine the morphology of a radio
source?  That is, why does it become Type I or Type II in the FR
classification?  Alternatively, why does it become Straight or Tailed,
the types introduced in this paper?

\item Does the morphology of a particular source change with time?
Does a source change from FR type II to I, or from Straight to Tailed?

\end{itemize}

In order to address these questions, we have undertaken a program of
study based on a new data set, the Ledlow-Owen sample \cite{LOrad} of
radio sources in clusters of galaxies.  
We expected that the evolution of a source should
be determined by such factors as its jet power, age, parent galaxy
size, and the density or pressure of the surrounding medium.  To this
end, we gathered radio, optical and X-ray data on the sources in the
sample, and have searched for trends which might answer the questions
above.  

In this process, we found that the Ledlow-Owen sources fall clearly
into two morphological classes. This was not what we expected.  
Nearly all of the sources in this sample would be called FR Type I,
based on their lack of outer hot spots, and their position in the
(power, magnitude) diagram (as presented by \cite{LObreak}).  Our work
with the images made it clear that the sources are not a uniform
sample, but rather divide nicely into two groups.  These groups are not well
described by the two well-known FR types.  
In fact, it now seems that the two FR types might more properly be
described as  ``classic Type II's'' (the rare sources with clear outer
hot spots) and ``all other sources'' (the more common sources, without
outer hot spots, and including the diversity of source types we find).

\section{The Data}

We worked with the Ledlow-Owen cluster sample (\cf\ \cite{LOrad} and
references 
therein).  The sample consists  of 219 radio sources above 10 mJy in
the inner region of nearby Abell clusters ($z < 0.09$), and a
representative sample of 48 sources above 200 mJy in more distant
clusters ($0.09 < z < 0.25$).    These data were supplemented by
images from the ROSAT all-sky survey  for a subset of the nearby
sample \cite{Voges}.  Details of the sample and our analysis of it are
given in \cite{Edata}. The data we have measured or compiled are:  
\begin{itemize}

\item Radio power at 1.4 GHz ($P_{1.4}$), \cite{LOrad}.

\item Radio images, also at 1.4 GHz, taken with the VLA \cite{LOrad}.

\item Optical magnitude of the parent galaxy, within an isophote at
24.5 mag asec$^{-2}$  in the R band ($M_{24.5}$); \cite{LOopt}. 

\item Radio flux sizes (the radius which includes $\gtw 90$\% of
the 1.4 GHz flux)  \cite{Edata}.

\item Source offset from the cluster X-ray peak (central or nearest
peak, depending on the cluster) \cite{Edata}.

\item Characteristic size of the X-ray cluster, measured in terms of
the radius which contains 50\% of the flux \cite{Edata}. 

\item Density of the cluster gas at the radio source position
(using simple deprojections of the X-ray data)  \cite{Edata}.  

\end{itemize}

Our initial goal was to use these data to address the fundamental
questions listed above.  This has not proved overly successful;
indeed, we found very few significant correlations in these
quantities.  We do find a slight
 correlation between radio power and radio size -- higher power
sources tend to be larger.  In general, however, statistical studies
of the full sample do not shed great light on the problems we set out
to study.

However, in this process we did find one very striking pattern. 
Figure 2 shows the full sample in the $(P_{1.4},M_{24.5})$ plane.  As
noted above, most of the sources fall in the ``Type I region'' as
defined by 
\cite{LObreak}.  Thus, we thought when we began that we were studying
Type I sources.  We found, however, a more complex situation.
 Nearly
all of the resolved sources fall neatly into two morphological
classes.  The existence of these two classes does not seem to have
been noticed before.  We believe it is a significant clue to
understanding the nature, and thus evolution, of radio galaxies in
general.

\section{The Two Classes}

To be specific:  197 of the sources have VLA images good enough to
estimate the source morphology.   (For the rest, 21 were unresolved,
and 48 were marginally resolved but too faint or too small to
allow us to determine the morphology).   Of the 197 well-resolved
sources, 188 can easily be put into one of our two
classes.\footnote{The remaining 9 include 6 amorphous, cluster-center
sources -- a different group interesting in their own right -- and 3
true ``oddballs''.}  Our criterion for classification is, {\em how far
does the jet propagate undisturbed?}.  Based on this, our two classes
are as follows.

{\bf Type A, or Straight, Sources.} In these sources, the jet retains
its identity all the way to the outer end of the source, where it may
or may not end in an outer bright spot. The end of the flow -- what
might be called a ``working surface'' -- is apparent in most of the
images; there is no 
continuation of the jet flow into an outer tail, as there is in Tailed
sources.  The inner jet is generally
embedded in a lobe or cocoon (although this may be faint and hard to
detect in some of the sources).  The jet can be more or less
collimated, depending on the source. These sources as a class are
not strongly bent.  The presence, or absence, of 
an outer hot spot is not critical to this class.  Sources with and
without external hot spots share all other characteristics.  Classical
FR Type II sources fall into this category, as do many sources which
would be 
called FR Type I.  About 1/3 of the sample (61/197) fall into this
category.   Thirteen of these have clear outer hot spots which would make
them classical FR Type II's. 

{\bf Type B, or Tailed, Sources.}  In these sources, the jet begins
narrow and well-collimated.  It soon  undergoes a dramatic
transition at an inner hot spot, where the flow becomes broader and
brighter.  High resolution images show complex structure within these
hot spots.  The flow does not disrupt, however; it continues on, past
the hot spot, into a radio tail.  In some sources the end of the tail
is seen, while the surface brightness of other sources fades too fast
to allow us to see the end.  This outer flow can sometimes  be seen to
have a two-part structure:  an inner channel embedded in an outer
lobe.  Sources which have been called Wide-Angle Tail or Narrow-Angle
Tail fall into this category.  We note, however, that while these
sources are often bent, they need not be.  Some Tailed sources display
the jet-hot spot transition without bending. About 2/3 of the sample
(127/197) fall into the Tailed category;  about 4/5 of these are
strongly bent.

\begin{figure}[htb]
\begin{minipage}{0.44\textwidth}
\rotatebox{180}{
 \resizebox{\textwidth}{!}{\includegraphics{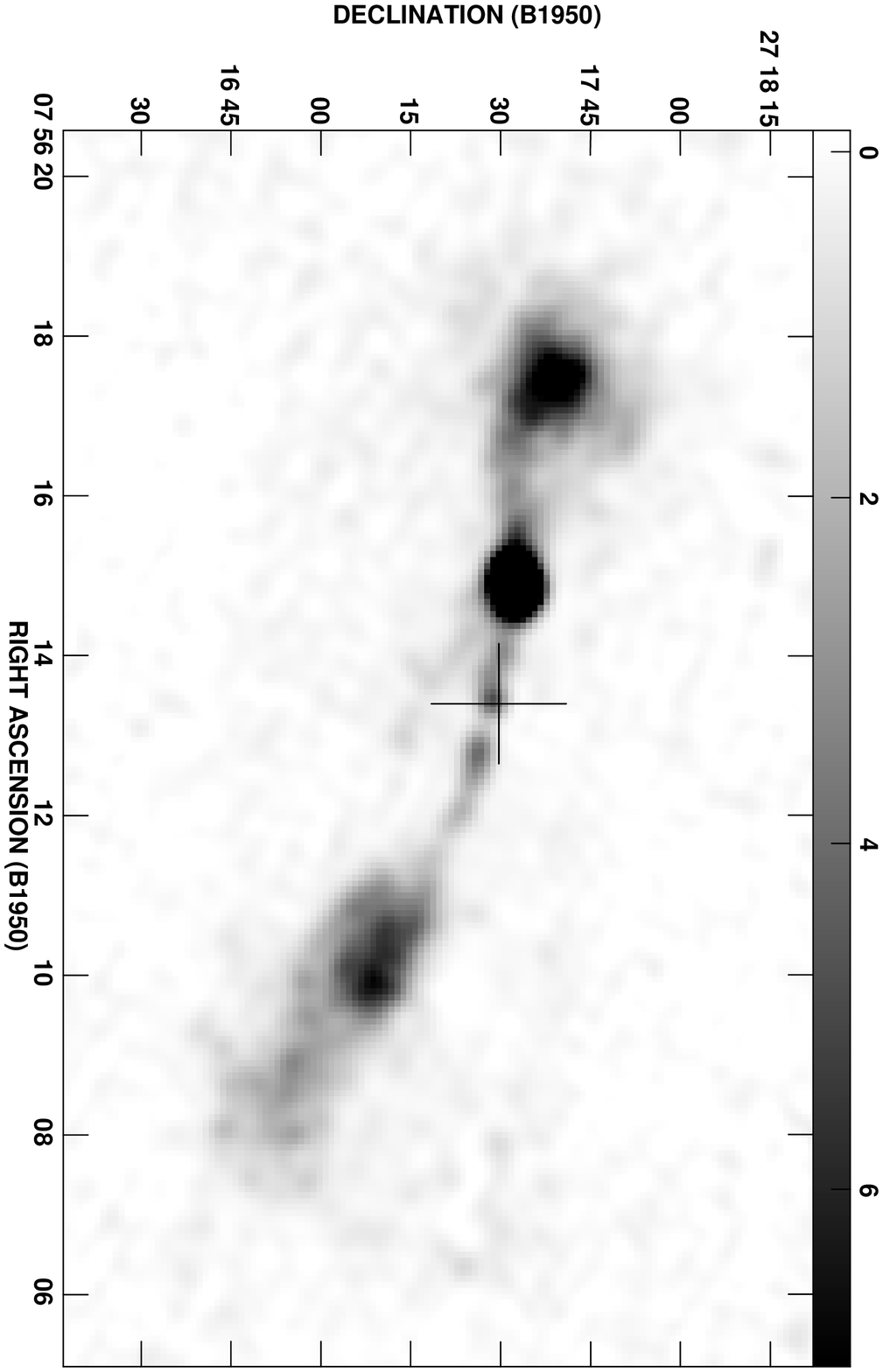}}}
\caption[]{The source 0756+272, an example of a tailed source. The
transition  from a narrow inner jet, through an interior hot spot, to
a broader tailed flow, is apparent. }
\end{minipage}
\hspace{0.04\textwidth}
\begin{minipage}{0.52\textwidth}
\rotatebox{180}{
\resizebox{\textwidth}{!}{\includegraphics{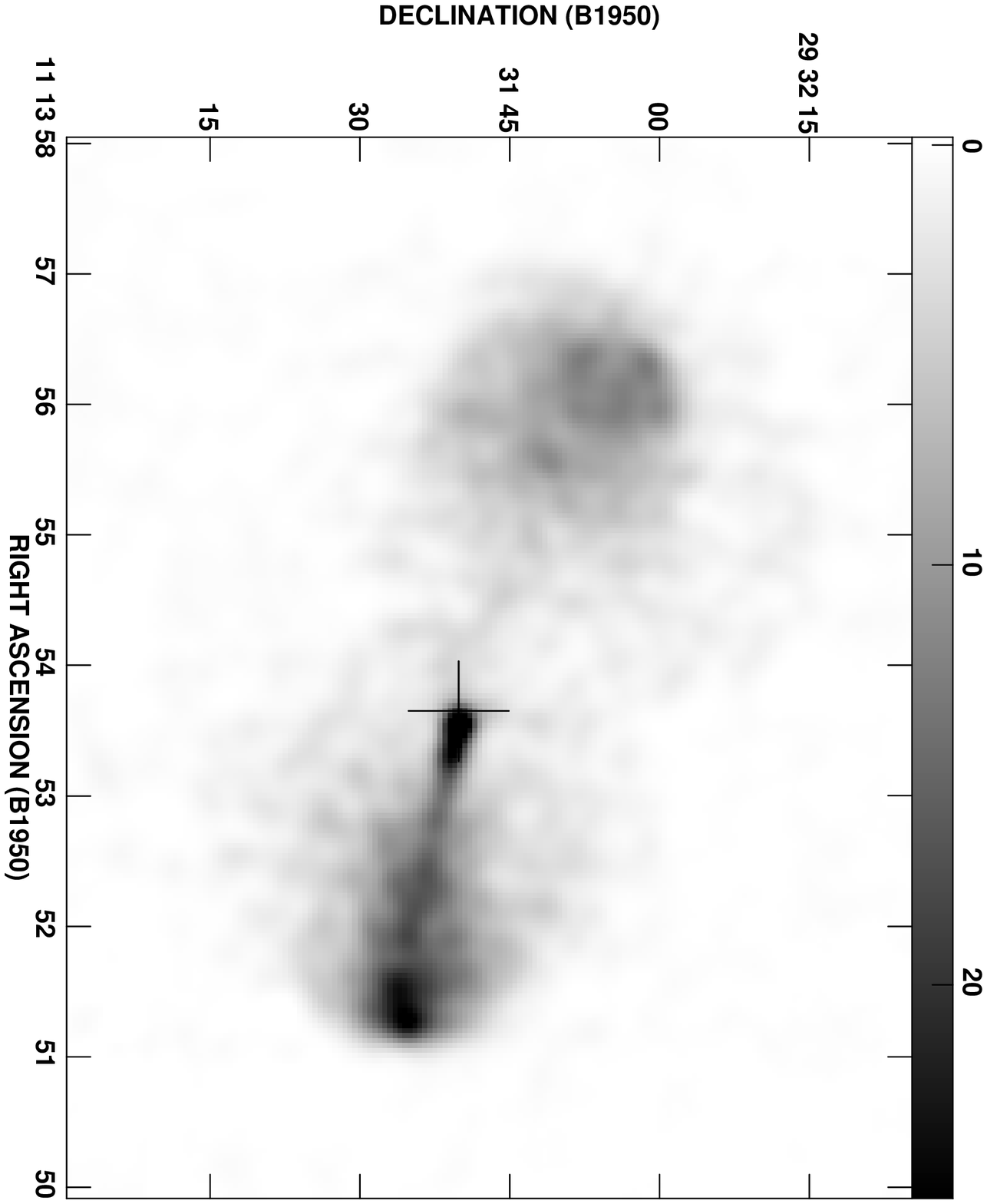}}}
\caption[]{The source 1113+295C, an example of a
straight source.  The inner jet can be seen within each lobe;  the jet
in the west lobe stays brighter and nearly forms a weak hot spot.}
\end{minipage}
\end{figure}

Figures 3 and 4 illustrate one example of each class;  more are
illustrated in \cite{Edata}, where we also present our  classification of
each source in the sample. 

We only have radio data at one frequency for this sample, so we cannot
draw general conclusions about the spectral properties of our two
classes.  However, other work provides hints.  Based on two-frequency
spectral maps,  Parma {\it etal} \cite{Parma} find that FR Type I sources
divide into two spectral types:  those in which the spectrum steepens
going away from the core, and those in which it flattens. Inspection
of the individual sources in their work suggests that their two
spectral types correlate with our two morphological types.  Based on
their smaller sample, we speculate that our Straight sources will turn
out to have spectra which flatten going away from the galactic core
(much as the spectra of FR Type II sources do).  We further speculate
that 
our Tailed sources will turn out to have spectra which steepen going
away from the core.  This speculation is also supported by the earlier 
work of O'Donoghue {\it etal} \cite{OOE}, who found that the radio
spectra of Wide-Angle Tail sources steepen going away from the core.

We believe the clear division of the sample into Straight and Tailed
sources is telling us something about the underlying physics of the
sources.  We were disappointed, however, to find that simple
statistical tests were inconclusive.  We might have expected, for
instance, that the sources would divide by radio size (perhaps one
type might be younger, or might evolve into the other?), or would
divide by radio power (a stronger jet might lead to different flow
dynamics?).  Alternatively, we might have expected the two types of
sources to sit in different cluster environments (perhaps high or low
ambient density might lead to different flow dynamics?).  We were
surprised to find that such simple ideas were not borne out. We 
looked for differences between the two classes in each of the  
quantities listed above, with little luck.  To zeroth order, Tailed
and Straight sources have the same radio power, same radio size, same
parent galaxies, and live in the same ranges of X-ray offsets and local gas
densities.

Thus, we have not been able to determine from our data a simple
observational factor, or factors, that  relate to a source becoming
Tailed or Straight.  We must go a step further and consider the
physics of the jet flow. In the next section we look to possible
physical mechanisms for the two dynamical classes.

\section{Dynamical Interpretation}

Our work with radio galaxies in clusters has uncovered a significant
point.  Two morphological (and therefore dynamical) classes of radio
galaxy exist. One class we call Straight (Type A).
In these sources 
the jet sits inside a lobe or cocoon, and continues mostly undisturbed
all the way from the core to the outer edge of the source.  The other
we call Tailed (Type B).  In these the jet flow changes dramatically
at an inner hot spot, but the flow continues on well past this point
into a tail (which also sits inside a larger cocoon).  We note that
our new groupings do not correlate particularly well with FR types I and
II.  All of our Tailed sources would be FR Type I;  however our
Straight sources contain both FR types.

What determines whether a source is Type A or Type B?  We were rather
disappointed to find that two classes do not separate in any of our
measured quantities.  They do not depend on jet power (unless the
conversion efficiency from jet to radiated power is quite different
for the two types).  They do not depend on local density or galaxy
mass (magnitude).  They are not a function of the source age (unless
the two types of sources reach the same linear size at very different
rates).  It follows that several simple arguments fail.  For instance,
we cannot say that Straight sources evolve into Tailed;  if that were
the case, we would have found that Tailed sources are larger than Straight.
Nor can we argue that something in the environment (such as ambient
pressure or density) gives rise to the difference;  there is no evidence
that the two source types are found in different regions of their clusters.

This leaves us only with the option of an internal variable.  That is,
some internal feature of the jet flow must determine whether the source 
becomes 
Tailed or Straight.  We suggest this is a matter of the stability
 of the flow.  Our hypothesis is as follows ({\it cf.} \cite{EH99} for
a more detailed discussion). 

\begin{itemize}

\item {\bf All young sources} must  start out with a ``straight''
configuration.  That is, they begin
with a jet propagating out into the local ambient
medium, and depositing material into a slower-moving cocoon which surrounds
the jet.  The young source will develop according to the usual arguments
based on momentum and energy conservation (for instance in
\cite{Scheuer}, \cite{ES}).
At some point in this evolution, internal factors decide whether the
source becomes Straight or Tailed.

\item {\bf Straight sources} result when the jet remains essentially
stable.  In these sources, the jet has not undergone dramatic
disruptions.   Rather, it
continues to propagate more or less undisturbed all the way to the end
of the source, reaching current lengths of several tens of kpc to 
a few hundred kpc.  At the end of the source, the jet impacts the
ambient medium, slows down and deposits material in a larger cocoon.
The lack of bent sources in this category requires that the 
undisurbed jet remain rigid, and thus not susceptible to deflection by
external forces. If the
jet remains highly supersonic, an outer shock (hot spot) will form.
If it is only subsonic or transonic, the outer jump will be smaller
and not necessarily radio bright.  Thus, classic FR Type II sources may
develop in this type of source, but need not do so.  

\item {\bf Tailed sources} result when the jet undergoes a strong instability.
The instability develops when the jet reaches a few  tens of kpc from
the galactic nucleus.  The location of the instability is apparent 
in images of these sources, where the jet enters the inner hot spots
and broader tails. 
When such an instability sets in and grows to large amplitude, it
distorts and disrupts the flow.  This results in side-to-side
``flapping'' of the ordered jet flow which  
can be seen in good images.  In addition, the instability appears to
saturate at some large amplitude.  The flow is not totally disrupted,
but is able to continue on to form a (quasi-stable) tail.  The images
show that such radio tails are often embedded in outer envelopes,
which we suspect act as ``cocoons'' for these flows.  When Tailed
sources are strongly bent, the bends occur at the hot spots:  the
change in flow properties there seems to make the flow more
susceptible to deflection.

\end{itemize}

\subsection{Stabilty of a Radio Jet}

  What governs the instability?  We consider the Kelvin-Helmholtz
instability as a concrete example.  This may very well be the most
important instability in jet dynamics; considering it as an example
allows us to take advantage of 
substantial work available in the literature\footnote{Current-driven
instabilities may also exist, of course.  We expect their behavior to
be qualitatively similar to that of the Kelvin-Helmholtz instability,
although with different governing parameters}.  Based on the work of
Hardee and his colleagues ({\it cf.} \cite{Phil}, \cite{RHDH}, and
references therein), three factors are critical to the jet's
stability: 

\begin{itemize}

\item {\bf The jet magnetic field}.  A strong parallel field
(compared to the jet's kinetic energy) will stabilize
the flow.  The Alfven Mach number (ratio of flow speed to Alfven
speed) is critical here.  As a general statement, we can say that
subAlfvenic flow is stable, while 
superAlfvenic is not.  This is modified by the jet density:  a jet
denser than its surroundings will remain stable to a somewhat higher
Alfven Mach number.  This is also extended by the field structure: a
strong helical field will be more effective at stablizing than a fully
parallel field.  

\item {\bf The jet density}.  Jets which are heavy relative to their 
surroundings are more stable.  This helps in two
ways.  (1) A heavy jet can remain stable even if somewhat
superAlfvenic. (2) Potentially disruptive perturbations on a linearly
unstable jet will grow more slowly if the jet is heavy.  We emphasize
that it is the {\it cocoon} density, not the ambient density, which is
relevant here.  

\item {\bf The jet speed}.  Instabilities in fast jets disrupt the
flow more slowly.  The distance over which an instability grows to
important levels, after it sets in, is proportional to the jet Lorentz
factor and to the jet magnetosonic Mach number (defined as the ratio
of jet speed to magnetosonic wave speed).  

\end{itemize}

A change in any one of these quantities can change a jet flow from
being nearly stable to being dramatically unstable.  For instance, it
may be that Straight sources result from more strongly magnetized
jets, and Tailed sources from jets with weaker fields.  Or, it may be
that Straight sources result from faster (or more relativistic) jets,
while Tailed sources are produced by slower (or subrelativistic)
jets.  Or, it may be that Straight sources result when the cocoon is
underdense relative to the jet, and Tailed sources when the (young)
cocoon is overdense.   We have not yet identified which of these
possibilities is the crucial one;  it may be that all three contribute
to the dynamics of real radio sources.

\subsection{Consequences of this picture}

How does this new picture impact arguments about source evolution and
global dynamics?  In particular, how do we address the questions
raised in \S 1?  The large-scale evolution of both Type A and Type B
sources must be governed by the outflow, following models originally
developed for FR Type II sources.   The linear size evolution 
will be determined by the jet thrust (balanced against the density of
the ambient medium).  The energy described by the jet power will be
partly deposited in the cocoon (the rest going to expansion work and
radiative losses).  Such models should be relevant to both classes of
sources we find, but must differ in application to the two classes.
For Straight sources, the thrust and power are those of the jet as it
emerges from the galactic core (perhaps on a  kpc scale).  For
Tailed sources, the thrust and power are those of the tail,
established  after the flow has reorganized itself at the inner hot 
spots.  

Finally, what can be said about the radio power?  
Predicting the evolution of radio power remains, as always, one of the
hardest questions.  We must remember that radio emission is a
sensitive, nonlinear, variable-gain tracer of the flow dynamics.
Synchrotron emission depends on a complex convolution of the magnetic
field strength and structure with the relativistic particle
distribution in space and in energy.   We remain cautious about
connecting dynamical models to emission models in any simple way.

\ack{This work was partially supported by NSF grant AST-9720263.}

\end{document}